\documentclass[12pt]{article}
\usepackage{latexsym}
\usepackage{amsmath,amsbsy,amssymb}

\newcommand{\wg}{\wedge}
\newcommand{\gam}{\gamma}
\newcommand{\Gam}{\Gamma}

\newcommand{\ddg}{\ddagger}
\newcommand{\tl}{\tilde}
\newcommand{\ul}{\underline}

\newcommand{\be}{\begin{equation}}
\newcommand{\bear}{\begin{eqnarray}}
\newcommand{\ear}{\end{eqnarray}}
\newcommand{\ee}{\end{equation}}
\newcommand{\lbl}{\label}
\newcommand{\bi}{\bibitem}
\newcommand{\ci}{\cite}

\newcommand{\vs}{\vspace}

\begin{document}

\

\baselineskip .7cm 

\vs{27mm}

\begin{center}

{\LARGE \bf A Novel View on the Physical Origin of E$_8$}

\vs{3mm}

Matej Pav\v si\v c

Jo\v zef Stefan Institute, Jamova 39,
1000 Ljubljana, Slovenia

e-mail: matej.pavsic@ijs.si

\vs{6mm}

{\bf Abstract}

\end{center}

We consider a straightforward extension of the 4-dimensional spacetime $M_4$
to the space of extended events associated with strings/branes, corresponding
to points, lines, areas, 3-volumes, and 4-volumes in $M_4$. All those objects
can be elegantly represented by the Clifford numbers $X\equiv x^A \gamma_A \equiv
x^{a_1 ...a_r} \gamma_{a_1 ...a_r},~r=0,1,2,3,4$.
This leads to the concept of the so-called Clifford
space ${\cal C}$,
a 16-dimensional manifold whose tangent space at every point is the Clifford algebra
${\cal C \ell }(1,3)$. The latter space besides an algebra is also a vector
space whose elements
can be rotated into each other in two ways: (i) either by the action of
the rotation matrices of SO(8,8) on the components $x^A$ or (ii) by the
left and right action of the Clifford numbers $R=$exp$\,[\alpha^A \gam_A]$ and
$S=$exp$\,[\beta^A \gam_A]$ on $X$. In the latter case, one does not
recover all possible rotations of the group SO(8,8). This
discrepancy between the transformations (i) and (ii) suggests that one should
replace the tangent space ${\cal C \ell }(1,3)$ with a vector space
$V_{8,8}$ whose basis elements are generators of the Clifford
algebra ${\cal C \ell }(8,8)$, which contains the Lie algebra
of the exceptional group E$_8$ as a subspace. E$_8$ thus arises
from the fact that, just as in the spacetime $M_4$ there are  $r$-volumes
generated by the tangent vectors of the spacetime, there are $R$-volumes,
$R=0,1,2,3,...,16$, in the Clifford space ${\cal C}$, generated by the tangent vectors
of ${\cal C}$.

\section{Introduction}

The unification of fundamental interactions has turned out to be an 
elusive goal. Amongst the numerous approaches that have been
pursued, Kaluza-Klein and string/brane theories are of
particular importance. While Kaluza-Klein theories\,\ci{Kaluza} introduce
extra dimensions into the game, string/brane theories\ci{strings}
incorporate the extended nature of physical objects. Although
very promising, these theories have encountered serious
problems that have remained unresolved, and, so far, they
have not led to a realistic model. In order to develop such a model, one has
to find a grand unification group, e.g., SU(5), SO(10),
E$_6$ or E$_8$, which fits into a more fundamental theory
that also incorporates the Poincar\'e group of spacetime.
For the time being, a lot of research is devoted to explaining
the properties of these groups regardless of their
possible deeper origin in physics.

Last year, Garrett Lisi \ci{Lisi} proposed that one could unify
all the fields of the Standard model and gravity by means
of the exceptional group E$_8$. Lisi's approach differs
from others \ci{e8} in attempting to explain generations in terms
of triality. It has attracted considerable
attention and criticism. Regardless of
whether the particular model due to Lisi will
turn out to be correct or not, it would be desirable
to find a possible physical basis for the group E$_8$.
In other words, we would like to know whether there
exists a particular physical arrangement that leads
to E$_8$. Is it due to the existence of higher-
dimensional dynamics? We will show that we do not
need extra dimensions of spacetime in order to
arrive at E$_8$. We can remain in the 4-dimensional spacetime
$M_4$ and consider not only events (points) but
also oriented lines, areas, 3-volumes and 4-volumes
in $M_4$. All these objects can be elegantly
represented by the Clifford numbers $X^A \gam_A
\equiv X^{a_1 a_2 ... a_r} \gam_{a_1 a_2 ... a_r}$, where $r=0,1,2,3,4$ and
$a=0,1,2,3$. This leads to the concept of the
so-called {\it Clifford space}\,\ci{Pezzaglia}--\ci{PavsicArena},
a 16-dimensional manifold whose tangent space is
the Clifford algebra\footnote{
In the introduction, we used a simplified notation
for various spaces, namely $M_4,~V_4,~ {\cal C \ell}(4),~
V_{16}$ and ${\cal C \ell}(16)$, even though Minkowski space
$M_4$ has signature (1,3) and should be written
more precisely as $M_{1,3}$, $V_4$ as $V_{1,3}$,
${\cal C \ell}(4)$ as ${\cal C \ell}(1,3)$, $V_{16}$ as
$V_{8,8}$ and ${\cal C \ell}(16)$ as ${\cal C \ell}(8,8)$.}
${\cal C \ell} (4)$.

In these papers it has been assumed that the arena for physics is
Clifford space ${\cal C}$ and that physical
objects are described by Clifford algebra valued
fields on ${\cal C}$. At every point $X \in {\cal C}$,
a field $\Phi$ can be transformed according to\,\ci{PavsicKaluza,PavsicKaluzaLong}
\be
    \Phi \rightarrow \Phi' = {\rm R} \Phi {\rm S}
\lbl{1.1}
\ee
where R and S are in general two different Clifford
numbers\,\ci{PavsicKaluza,PavsicKaluzaLong} to be specified later. For the
time being, let us only say that the transformation
(\ref{1.1}) in general reshuffles the grade $r$
 of the components $\Phi = \phi^A \gam_A$. Its
 action on $\phi^A$ or $\gam_A$ can be just a
 {\it rotation}. Altogether (if $D=2^n$ is the dimension of
 ${\cal C \ell}(n)$)  there are $D(D-1)/2$
 independent rotations, 120 if $n=4$, but they cannot
 all be generated by the 16 independent $\gam_A$ which
 close ${\cal C \ell}(4)$. However, we can perform
 rotations in ${\cal C \ell}(4)$ by means of the
 rotation matrices acting on $\phi^A$ according to
 \be
      \phi'^A = {L^A}_B \phi^B
\lbl{1.2}
\ee
The existence of 120 rotational degrees of freedom
implies the presence of extra degrees of freedom
besides those spanned by the basis $\{\gam_A \}$.

At this point, we will employ the fact that the
Clifford algebra ${\cal C \ell}(4)$ is a vector space.
Eq.\,(\ref{1.1}) performs only 32 independent
rotations in the vector space (16 due to 
the left and 16 due to the right transformations).
Therefore, instead of the basis $\{\gam_A \}$, we may
use an alternative basis $\{\Gam_A \}$, $A=1,2,...,16$,
whose elements $\Gam_A$ generate the Clifford
algebra ${\cal C \ell}(16)$. The new basis elements
$\Gam_A$ span a 16-dimensional space $V_{16}$ which,
as a vector space, is isomorphic to ${\cal C \ell}(4)$.
Rotations in $V_{16}$ are generated by bivectors
$\Gam_{AB} \equiv {\cal C \ell}(16)$; they act on
a vector $\phi^A \Gam_A \in V_{16}$ and belong to
the Lie algebra of the group SO(16). The latter
group has not only a vector representation but also
spinor representations. In our set up 
spinors come from the Clifford
algebra ${\cal C \ell}(16)$, namely, from its left
(right) minimal ideals. The Lie algebra so(16),
together with the space ${\rm S}_{16}^+$ of real,
positive chiral ${\cal C \ell}(16)$ spinors, forms
the Lie algebra ${\rm e}_8$ of ${\rm E}_8$. 

So, we have found that e$_8$ arises from the spacetime
$M_4$ itself, from which one can construct a Clifford algebra.
The latter algebra, as a vector space, is isomorphic
to $V_{16}$, and its basis vectors $\Gam_A$ can generate the
Clifford algebra ${\cal C \ell}(16)$, whose subspace
is e$_8$:
\be
   V_4 \rightarrow {\cal C \ell}(4) \sim V_{16} 
   \rightarrow {\cal C \ell}(16) \supset e_8
\lbl{1.3}
\ee

In the following, we will explore this procedure in more
detail.

\section{Generalizing spacetime}

Special and general relativity are valid in the spacetime manifold
$M_4$. Choosing a point ${\cal P} \in M_4$, the tangent
space at that point is a vector space 
$V_4 \equiv T_{\cal P} (M_4)$, whose basis vectors can be
taken to be generators $\gam_a,~a=0,1,2,3$, of the Clifford
algebra, satisfying
\be
   \gam_a \cdot \gam_b \equiv \mbox{$\frac{1}{2}$}
   (\gam_a \gam_b + \gam_b \gam_a) = \eta_{ab}
\lbl{2.1}
\ee
where $\eta_{ab}= {\rm diag}\, (1,-1,-1,-1)$ is the Minkowski
metric tensor.

While the symmetrized product (\ref{2.1}) gives the metric,
the antisymmetrized (wedge) product
\be
   \gam_a \wg \gam_b \equiv \mbox{ $\frac{1}{2}$}
   [\gam_a,\gam_b] = \mbox{ $\frac{1}{2}$}
   (\gam_a \gam_b - \gam_b \gam_a)
\lbl{2.2}
\ee
gives a bivector. In general, the antisymmetrized product
of $r$ vectors
\be
   \gam_{a_1 a_2 ... a_r} \equiv \gam_{a_1} \wg \gam_{a_2}
   \wg ... \wg \gam_{a_r} \equiv \frac{1}{r!}\,
   [\gam_{a_1},\gam_{a_2},...,\gam_{a_r}] \equiv \gam_A
\lbl{2.2a}
\ee
gives an $r$-vector. The objects $\gam_{a_1 a_2 ... a_r}
\equiv \gam_A$ form a basis of the Clifford algebra
${\cal C \ell}(1,3)$, whose elements are Clifford numbers
called {\it polyvectors} and which are given by a superposition
\be
    X = X^{a_1 ... a_r}\gam_{a_1 ... a_r} \equiv X^A \gam_A
\lbl{2.3}
\ee
We take $a_1 < a_2 < ... < a_r$ in order to avoid multiple
occurrences of the same term.

The components $X^{a_1 ... a_r} \equiv X^A$ are holographic 
projections of oriented $r$-areas onto the coordinate
$r$-planes, spanned by $\gam_{a_1 ... a_r} \equiv \gam_A$.

In the spacetime manifold $M_4$, in which we choose a point 
${\cal P}_0$ as the origin, the vectors $x = x^a
\gam_a \in V_4 ({\cal P}_0) = T_{{\cal P}_0}(M_4)$
are in one-to-one correspondence with other points
${\cal P} \in M_4$, at least within a region $B\subset M_4$
around ${\cal P}_0$. 

Similarly, the polyvectors $X=X^A \gam_A \in {\cal C \ell}(1,3)$
are in one-to-one correspondence with the points ${\cal E}$ of a
16-dimensional manifold ${\cal C}$, called a {\it Clifford space},
in which we choose a point ${\cal E}_0$ as the "origin,"
such that $\gam_A \in T_{{\cal E}_0} ({\cal C})$ are tangent
polyvectors at ${\cal E}_0$. In other words, vectors of the
tangent space at ${\cal E}_0 \in {\cal C}$ can be mapped
into the points ${\cal E}\in {\cal C}$ of a region
around ${\cal E}_0$. A tangent space to a point
${\cal E} \in {\cal C}$ is the vector space defined by the Clifford algebra
${\cal C \ell}(1,3)$.

The metric is given by the scalar product
\be
G_{AB} = \gam_A^\ddg * \gam_B \equiv \langle
    \gam_A^\ddg  \gam_B \rangle_0 
\lbl{2.4}
\ee
Here $\ddg$ reverses the order of the 1-vectors ---for example,    
    $ (\gam_{a_1} \gam_{a_2} .... \gam_{a_r})^\ddg =
     \gam_{a_r} ... \gam_{a_2} \gam_{a_1}$ --- 
and is necessary for the consistent raising and lowering
of indices\,\ci{PavsicMaxwell,PavsicKaluzaLong}.
With the definition (\ref{2.4}), the signature of ${\cal C}$ is
$(8,8)$.

Since a Clifford space is such a straightforward
generalization of spacetime, its has been assumed
\ci{Pezzaglia}--\ci{PavsicArena} that physics also
has to be generalized. According to such a novel theory,
the arena for physics is not the spacetime $M_4$
but the Clifford space ${\cal C}$ (also called ${\cal C}$-space).
The fundamental building blocks of physical space are then
not points of spacetime but extended $r$-dimensional
cells ($r$-cells), or extended events described by
$r$-vectors. For definiteness, we may imagine that
extended events are associated with $p$-branes in
the quenched mini\-superspace description \ci{AuriliaQuenched,
AuriliaFuzzy}, but, in principle, extended events can
be related to whichever fundamental extended objects that can
be described by $r$-vectors.

\section{Group SO(8,8) acting in Clifford space}
\subsection{Left and right transformations of
a polyvector}

A polyvector $X\in {\cal C \ell}(1,3)$ can be transformed
into another polyvector $X' \in {\cal C \ell}(1,3)$
by
\be
   X' = {\rm R}\, X \, {\rm S}
\lbl{3.1}
\ee
where R, S $\in {\cal C \ell}(1,3)$. Requiring that the
norm of $X$ be invariant
\be
   |X'|^2 \equiv {X'}^{\,\ddg}*X' \equiv \, \langle {X'}^{\,\ddg}
   X'\rangle_0\, = \, \langle{\rm S}^\ddg X^\ddg {\rm R}^\ddg {\rm R}
   X {\rm S}\rangle_0 = \langle X^\ddg X \rangle_0
\lbl{3.2}
\ee
we have
\be
   {\rm R}^\ddg {\rm R} =1 ~, ~~~~~~~~~~~{\rm S}^\ddg {\rm S}\,= 1
\lbl{3.3}
\ee

Writing
\be
    {\rm R} = {\rm e}^{\frac{1}{2}\alpha^A \gam_A} \, ,\qquad ~~ 
   {\rm S} = {\rm e}^{\frac{1}{2}\beta^A \gam_A}\, ,
\lbl{3.4}
\ee
we have from eq.\,(\ref{3.3}) that
\be
               \alpha^* \gam_A^\ddg = -\,\gam_A \alpha^A
\lbl{3.5}
\ee
where we assume that reversion also acts on the imaginary
unit $i$, which we consider a bivector $i=e_q e_p$ of the
phase space spanned by $e_q$ and $e_p$ \ci{PavsicBook}.
The parameters $\alpha^A,~\beta^A,~A=1,2,...,16$
are, therefore, either real or imaginary because, depending on
the grade of $\gam_A$, we have:
\be
    \gam_A^\ddg = \gam_A ~~~ {\rm if}~ A \in \{A\}_+\, ,~~~
   \gam_A^\ddg = -\, \gam_A ~~~ {\rm if} ~A \in \{A\}_-
\lbl{3.6}
\ee

As an example, let us consider the case
\be
   {\rm R} = {\rm e}^{\frac{1}{2}\alpha \gam_1 \gam_2}=
   {\rm cos} \, \frac{\alpha}{2}+ \gam_1 \gam_2 \, 
   {\rm sin} \, \frac{\alpha}{2} \; , ~~~
  {\rm S} = {\rm e}^{\frac{1}{2}\beta \gam_1 \gam_2}=
   {\rm cos} \, \frac{\beta}{2}+ \gam_1 \gam_2 \, 
   {\rm sin} \, \frac{\beta}{2}
\lbl{3.7}
\ee
and examine how various Clifford numbers $X=X^C \gam_C$
transform under (\ref{3.1}).

(i) If $X= X^1 \gam_1 + X^2 \gam_2$, then the transformation
(\ref{3.1}) gives
\be
   X' = X^1 \left ( \gam_1 \, {\rm cos}\, \mbox{$\frac{\alpha-\beta}{2}$}\,+
   \gam_2 \, {\rm sin}\, \mbox{$\frac{\alpha-\beta}{2}$} \right ) +
   X^2 \left ( -\gam_1 \, {\rm sin}\, \mbox{$\frac{\alpha-\beta}{2}$}\,+
   \gam_2 \, {\rm cos}\, \mbox{$\frac{\alpha-\beta}{2}$} \right )
\lbl{3.8}
\ee

(ii) If $X= X^3 \gam_3 + X^{123}\, \gam_{123}$, then
\be   
    X' = X^3 \left ( \gam_3 \, {\rm cos}\, \mbox{$\frac{\alpha+\beta}{2}$}\,+
   \gam_{123} \, {\rm sin}\, \mbox{$\frac{\alpha+\beta}{2}$} \right ) +
   X^{123} \left ( -\gam_{2} \, {\rm sin}\, \mbox{$\frac{\alpha+\beta}{2}$}\,+
   \gam_{123} \, {\rm cos}\, \mbox{$\frac{\alpha+\beta}{2}$} \right )
\lbl{3.9}
\ee   

(iii) If $X = s\, {\underline 1} + X^{12}\, \gam_{12}$, then
\be
     X' = s \left ( {\underline 1} \, {\rm cos}\, \mbox{$\frac{\alpha+\beta}{2}$}\,+
   \gam_{12} \, {\rm sin}\, \mbox{$\frac{\alpha+\beta}{2}$} \right ) +
   X^2 \left ( -{\underline 1} \, {\rm sin}\, \mbox{$\frac{\alpha+\beta}{2}$}\,+
   \gam_2 \, {\rm cos}\, \mbox{$\frac{\alpha+\beta}{2}$} \right )
\lbl{3.10}
\ee

(iv) If $X= {\tl X}^1 \, \gam_5 \gam_1 + {\tl X}^2 \, \gam_5 \gam_2$,
then
 \be
   X' = {\tl X}^1 \left ( \gam_5 \gam_1 \, {\rm cos}\,
   \mbox{$\frac{\alpha-\beta}{2}$}\,+
   \gam_5 \gam_2 \, {\rm sin}\, \mbox{$\frac{\alpha-\beta}{2}$} \right ) +
   {\tl X}^2 \left ( -\gam_5 \gam_1 \, {\rm sin}\,
    \mbox{$\frac{\alpha-\beta}{2}$}\,+
  \gam_5 \gam_2 \, {\rm cos}\, \mbox{$\frac{\alpha-\beta}{2}$} \right )
\lbl{3.10a}
\ee    

The usual rotations of vectors or pseudovectors are reproduced
if the angle $\beta$ for the right transformation is
equal to minus the angle $\alpha$ for the left transformation, i.e.,
if 
\be
\beta = -\alpha
\lbl{3.11}
\ee
Then all other transformations which mix the grade vanish.

Condition (\ref{3.11}) is a prerequisite for the consistency of
ordinary rotations and Lorentz transformations. If we release the
condition (\ref{3.11}) and allow for arbitrary independent
$\alpha$ and $\beta$, in particular $\alpha=\beta$, then
we go outside the Lorentz group in $M_4$ and arrive at
rotations in Clifford space.

In the example (\ref{3.7}), the generator $\gam_1 \gam_2$
changes sign under reversion, so the parameters $\alpha$ and
$\beta$ are real. If we consider, for example, the
rotations generated by $\gam_1$, then, since $\gam_1^\ddg =
\gam_1$, the transformation parameter must be imaginary,
which we write explicitly as
  \be
   {\rm R} = {\rm e}^{\frac{1}{2}\,i\, \alpha \gam_1}=
   {\rm ch} \, \frac{\alpha}{2}+ i\,  \gam_1 \, 
   {\rm sh} \, \frac{\alpha}{2} \,,  ~~~
  {\rm S} = {\rm e}^{\frac{1}{2}\, i\, \beta \gam_1}=
   {\rm ch} \, \frac{\beta}{2}+ i \,\gam_1 \, 
   {\rm sh} \, \frac{\beta}{2}
\lbl{3.11a}
\ee
Taking $X= s {\ul 1}$ or $X = X^1 \gam_1$, we have
\be
    {\rm R}\, s {\ul 1}\, {\rm S} = s \left ({\ul 1} \, {\rm ch}\, 
    \mbox{$\frac{\alpha+\beta}{2}$} + i \, \gam_1 \, {\rm sh}\,
    \mbox{$\frac{\alpha+\beta}{2}$}\right )
\lbl{3.11b}
\ee 
\be
    {\rm R}X^1 \, \gam_1 {\ul 1}\, {\rm S} = X^1 
    \left ( -i \, {\ul 1} \, {\rm ch}\, 
    \mbox{$\frac{\alpha+\beta}{2}$} +  \gam_1 \, {\rm sh}\,
    \mbox{$\frac{\alpha+\beta}{2}$}\right )
\lbl{3.11c}
\ee   
We see that $i \gam_1$ generates rotations in the plane
$(i\, {\ul 1}, \gam_1)$ or $({\ul 1}, i\gam_1)$.

For infinitesimal transformations we have
\be
  X' = {\rm e}^{\frac{1}{2}\alpha^A \gam_A }X^C \gam_C\,
       {\rm e}^{\frac{1}{2}\beta^B \gam_B } \approx
       1 + \delta X
\lbl{3.12}
\ee
where
\bear
\delta X &=& \delta X^C \gam_C = \mbox{ $\frac{1}{2}$}
   (\alpha^A \gam_A \gam_C + \beta^A \gam_C \gam_A) X^C \nonumber \\
  &=& \mbox{$\frac{\alpha^A + \beta^A}{4}$} \, \{\gam_A,\gam_C \}X^C
   \,+ \, \mbox{$\frac{\alpha^A - \beta^A}{4}$} \, [\gam_A,\gam_C] X^C \nonumber \\
   &=& \mbox{$\frac{\alpha^A + \beta^A}{4}$} \,f_{AC}^D \gam_D X^C
   \,+ \, \mbox{$\frac{\alpha^A - \beta^A}{4}$} \, C_{AC}^D \gam_D X^C
\lbl{3.13}
\ear
where $\{ \gam_A,\gam_C\}= f_{AC}^D \, \gam_D$ and $[\gam_A,\gam_C]=
 C_{AC}^D \,\gam_D$. Introducing
 \be
    {\epsilon^D}_C =\mbox{$\frac{\alpha^A + \beta^A}{4}$} \,f_{AC}^D \gam_D
    \,+\, \mbox{$\frac{\alpha^A - \beta^A}{4}$} \, C_{AC}^D \gam_D
\lbl{3.14}
\ee
we find
\be
   \delta X^D = {\epsilon^D}_C\, X^C
\lbl{3.15}
\ee
which denotes infinitesimal rotations of the ${\cal C}$-space
coordinates $X^C$ if ${\epsilon^D}_C$ is antisymmetric, and other
linear transformations otherwise. Eq.\,(\ref{3.14}) implies that
there are only 32 independent components of ${\epsilon^D}_C$,
of which 16 are due to $\alpha^A$, and 16 to $\beta^A$.

A single generator, e.g., $\gam_{12}$ in examples
(\ref{3.7})--(\ref{3.10a}), generates rotations not in a single
plane but in several planes of ${\cal C}$-space at once.
An additional complication is that, in
order to satisfy the requirement (\ref{3.3}) for the
invariance of the norm, some transformation parameters
must be real, and some must be imaginary. Therefore, some rotations
are within ${\cal C \ell}(1,3)$ while others are between
${\cal C \ell}(1,3)$ and  $i\,{\cal C \ell}(1,3)$, as
illustrated in (\ref{3.11a})--(\ref{3.11c}). Our space
is then enlarged to 
$\mathbb{C}\otimes {\cal C \ell}(1,3)$.

So, we have extended real polyvectors to complex ones.
In quantized theories, physical particles are described by
fields that, in general, are complex valued. In our case,
these fields are $\Phi = \phi^A \gam_A$, whose values are
in a complexified Clifford algebra. The quadratic form is
\be
   \Phi^\ddg *\Phi = {\phi^\ddg}^A \phi^B G_{AB} =
   \phi_{(R)}^A \phi_{(R)}^B \, G_{AB} +
    \phi_{(I)}^A \phi_{(I)}^B \, G_{AB}
\lbl{3.16}
\ee
where $G_{AB}$ is defined according to eq.\,(\ref{2.4}).
    
The real components $\phi_{(R)}^A$ can be transformed amongst
themselves by rotations of the group SO(8,8), and the
same is true for the imaginary components:
\be
   \phi_{(R)}^{'A} = {L^A}_B \phi_{(R)}^B \, , \quad ~~~~
    \phi_{(I)}^{'A} = {L^A}_B \phi_{(I)}^B
\lbl{3.18}
\ee
There are also mixed transformations between
$\phi_{(R)}^{A}$ and $\phi_{(I)}^{A}$. The Lie algebra
of SO(8,8) has 120 independent generators. Altogether,
the group of transformations that preserve the quadratic
form (\ref{3.16}) has 120 dimensions due to rotations
of $\phi_{(R)}^{A}$, 120 dimensions due to rotations
of $\phi_{(I)}^{A}$, and $16 \times 16 =256$ dimensions
belonging to mixed transformations between $\phi_{(R)}^{A}$
and $\phi_{(I)}^{A}$, which sums to 496. This, of course,
is in agreement with the dimension $32 \times 31/2 = 496$
of the SO(16,16) acting on $(\phi_{(R)}^{A},\phi_{(I)}^{A})$.

\subsection{Vector space $V_{8,8}$}

We have seen that, amongst left and right transformations
of a real polyvector $X$, there are norm-preserving
transformations, i.e., rotations that preserve or
change the grade of the elements $\gam_A \in
{\cal C}(1,3)$. Therefore, besides the rotations in the
planes $(\gam_\mu , \gam_\nu )$ and 
$(\gam_5 \gam_\mu , \gam_5 \gam_\nu )$, we also have
rotations in the planes $(\gam_\mu , \gam_{\rho \sigma})$
spanned by vectors and bivectors, etc. Such grade
-mixing rotations generated by (\ref{3.1}) do not act in
a single plane but in several planes at once, as
illustrated in examples (\ref{3.7})--(\ref{3.10a}).
Moreover, there are planes in ${\cal C \ell}(1,3)$ in
which rotations cannot be generated by a transformation
(\ref{3.1}). For instance, a rotation in the plane
$({\ul 1},\gam_1)$ does not occur in (\ref{3.1}).
Instead, one has a rotation in the plane
$({\ul 1},i\gam_1)$, which is not in ${\cal C \ell}(1,3)$,
but in $\mathbb{C}\otimes {\cal C \ell}(1,3)$.

A basis $\gam_A \in {\cal C \ell}(1,3)$ cannot generate
all possible rotations of the group SO(8,8)
acting within ${\cal C \ell}(1,3)$. However, we
can construct the SO(8,8) rotations directly, without
using $\gam_A$ as generators:
\be
   X' = X'^A \gam_A = {L^A}_B X^B \gam_A = X^A \gam'_A
\lbl{3.19}
\ee
where
\be
     X'^A =  {L^A}_B X^B \, , ~~~~~~\gam'_A = {L^B}_A \gam_B
\lbl{3.20}
\ee
So, either the components $X^A$ or the basis (poly)vectors $\gam_A$
can be rotated by a matrix ${L^A}_B$ of SO(8,8).

If we consider ${\cal C \ell}(1,3)$ as a vector space
$V_{8,8}$ and forget about the algebra, then,
instead of $\gam_A \in {\cal C \ell}(1,3)$, we can use any
other elements that span $V_{8,8}$.
Let us choose as our basis vectors the generators
$\Gam_A,~A=1,2,...,16$ of ${\cal C \ell}(8,8)$. The generators of
rotations in $V_{8,8}$ are then the 120 independent
bivectors $\Gam_{AB} = \Gam_A \wg \Gam_B \in {\rm so}(8,8) \subset
{\cal C \ell}(8,8)$.

Our space is thus the 16-dimensional vector space $V_{8,8}$,
spanned by the vectors
$\Gam_A \in {\cal C \ell}(8,8)$. An element $X=X^A \Gam_A
\in V_{8,8}$ is rotated according to
\be
  X' = {\rm e}^{\frac{1}{4}\alpha^{AB} \Gam_{AB}}\, X
  {\rm e}^{-\frac{1}{4}\alpha^{AB} \Gam_{AB}}
\lbl{3.21}
\ee
By also taking into account the ${\cal C \ell}(8,8)$
spinors, we arrive at the Lie algebra e$_8$ as a subspace
of ${\cal C \ell}(8,8)$. 

\section{Vector space $V_8^{(1)} \oplus V_8^{(2)}$}

Now, we will describe an alternative procedure that also leads
from ${\cal C \ell}(1,3)$ to e$_8$. We can split
${\cal C \ell}(1,3)\sim V_{8,8}$ into two subspaces
$V_{2,6}$ and $V_{6,2}$ spanned by the following basis
elements:
\be
   V_{2,6}\;:~~~\{\gam_{A_1}\} = \{1,\gam_\mu, \gam_{0r}\}\; ,
  ~~ A_1 = 1,2,...,8 \nonumber
\ee
 \be
   V_{6,2}\;:~~~\{\gam_{A_2}\} = \{\gam_5,\gam_5 \gam_\mu, 
   \gam_5 \gam_{0r}\}\;,
   ~~A_2 = 9,10,...,16
\lbl{4.1}
\ee
where the three elements $\gam_5 \gam_{0r}$, $r=1,2,3$, correspond to
the three elements $\gam_{rs}$. Every element
$\Phi\in {\cal C \ell}(1,3)$ can thus be split into two parts
in the following way:
\be
   \Phi = \phi^A \gam_A = 
   \phi^{A_1} \gam_{A_1} + \phi^{A_2} \gam_{A_2}
\lbl{4.2}
\ee
For convenience, we perform the split ${\cal C \ell}(1,3)
\sim V_{8,8} = V_{2,6} \oplus V_{6,2}$. Alternative splits,
such as  $V_{8,8} = V_{0,8} \oplus V_{8,0}=
V_{1,7} \oplus V_{7,1}$, etc., are also possible in principle.
For simplicity, we will from now on use the symbol $V_8$ for all
these possible spaces and consider the generic
split
\be
    {\cal C \ell}(1,3) \sim V_{8,8} =
    V_8^{(1)} \oplus  V_8^{(2)}
\lbl{4.3}
\ee

The basis elements $\gam_A \equiv \gam_{a1...a_r}$, $r=0,1,2,3,4$,
given by eq.\,(\ref{2.2a}), are $r$-vectors of 
${\cal C \ell}(1,3)$. Instead of a basis consisting of
16 $r$-vectors, we can consider a basis that consists of
16 independent spinors $\xi_{\tl A} \equiv \xi_{\alpha i}$,
$\alpha = 1,2,3,4$, $i=1,2,3,4$, that span four independent
left minimal ideals of ${\cal C \ell}(1,3)$ \ci{Teitler,PavsicKaluzaLong}.
Then, $\Phi$ can be expanded according to
\be
  \Phi = \psi^{\tl A} \xi_{\tl A} = \psi^{\alpha 1} \xi_{\alpha 1} +
  \psi^{\alpha 2} \xi_{\alpha 2} +\psi^{\alpha 3} \xi_{\alpha 3}+
  \psi^{\alpha 4} \xi_{\alpha 4}    
\lbl{4.4}
\ee
Four linearly independent spinors of one minimal left
ideal form a 4-dimensional representation of ${\cal C \ell}(1,3)$.
In terms of $\xi_{\alpha i}$, for a fixed $i$, say $i=1$,
every element of ${\cal C \ell}(1,3)$ can be represented
as a $4\times 4$ matrix. In terms of all 16 basis elements
$\xi_{\tl A}$, every element $\Phi \in {\cal C \ell}(1,3)$
can be represented as a $16 \times 16$ matrix according to
\ci{PavsicKaluzaLong}
\be
    \langle {\xi^{\tl A}}^\ddg \Phi \xi_{\tl B} \rangle_S =
    {(\Phi)^{\tl A}}_{\tl B} = {\delta^i}_j \otimes
    {(\Phi)^\alpha}_\beta
\lbl{4.5}
\ee
where $\alpha,~\beta$ are the 4-spinor indices of a given
left minimal ideal and the bracket with subscript $S$ denotes
the normalized scalar part\,\ci{PavsicKaluzaLong}.
The above formula explicitly shows
that the 16-dimensional representation of ${\cal C \ell}(1,3)$ is
reducible: the matrices are block-diagonal, with the blocks
being just $4 \times 4$ matrices. In general, instead of
16 spinors $\xi_{\tl A} \equiv \xi_{\alpha i}$, one can use some
other set of 16 linearly independent objects $q_A$,
each of which is a superposition of $\xi_{\tl A}$. A matrix
\be
   {(\Phi)^A}_B = \langle {q^{A}}^\ddg \Phi q_{B} \rangle_S
\lbl{4.5a}
\ee
then may not be block-diagonal, but, of course, it can be
transformed back into block-diagonal form. If $q_A = \gam_A$,
then $\langle {\gam^{A}}^\ddg \Phi \gam_{B} \rangle_S$ is
a representation of a Clifford number $\Phi$ in the
Clifford algebra basis $\gam_A$.

Let us now generalize from
{\it reducible} $16 \times 16$ matrices to
{\it arbitrary} $16 \times 16$ matrices. An arbitrary
$16 \times 16$ matrix can be expanded in terms of 256
linearly independent $16 \times 16$ matrices that represent
a basis of the Clifford algebra ${\cal C \ell}(8)$ (or one of 
its non-compact forms), where $256 = 2^8$ is the dimension
of ${\cal C \ell}(8)$ and $16= 2^{8/2}$ is the dimension of
a corresponding space that forms an irreducible representation
of ${\cal C \ell}(8)$.

Let us now return to the split (\ref{4.3}). Instead of the
basis elements $\gam_{A_1},~ \gam_{A_2} \in {\cal C \ell}(1,3)$
that, according to (\ref{4.5}) and (\ref{4.5a}), can be
represented by {\it reducible} (block diagonal) $16 \times 16$
matrices, let us now consider new basis elements
$\Gam_{\bar A}^{(1)}$ and $\Gam_{\bar A}^{(2)}$, ${\bar A}=1,2,...,8$,
that are 1-vectors of ${\cal C \ell}(8)$, represented by
$16 \times 16$ matrices that, in general, are not reducible
to a block diagonal form (\ref{4.5}). Every element of
$V_8^{(1)}$ can now be expanded in terms of $\Gam_{\bar A}^{(1)}$,
and analogously for $V_8^{(2)}$ and $\Gam_{\bar A}^{(2)}$.

Besides the 1-vectors
$\Gam_{\bar A}^{(1)}\in V_8^{(1)} \subset {\cal C \ell}(V_8^{(1)})$
and 
$\Gam_{\bar A}^{(2)}\in V_8^{(2)} \subset {\cal C \ell}(V_8^{(2)})$,
which are the Clifford algebra generators of the two copies
of ${\cal C \ell}(8)$, let us consider also 2-vectors
$$\Gam_{{\bar A}{\bar B}}^{(1)} \in {\rm so}(V_8^{(1)})~~ {\rm and}
~~\Gam_{{\bar A}{\bar B}}^{(2)} \in {\rm so}(V_8^{(2)})$$
The latter objects are generators of rotation in
$V_8^{(1)}$ and $V_8^{(2)}$, respectively.

In addition to the rotations in each of the spaces
$V_8^{(1)}$, $V_8^{(2)}$, there are also transformations
that mix those two spaces, and they are given by the generators
$\Gam_{\bar A}^{(1)} \otimes\Gam_{\bar A}^{(2)}$. Altogether,
there are 28 generators of ${\rm so}(V_8^{(1)})$, 28 generators
of ${\rm so}(V_8^{(2)})$, and $8 \times 8 =64$ generators
$\Gam_{\bar A}^{(1)} \otimes\Gam_{\bar A}^{(2)}$, which sums
to a total of 120 generators. This is the dimension of
so(16) that can be decomposed according to \ci{BaezOctonions,CastroE8}
\be
    {\rm so} (16) =  {\rm so}(V_8^{(1)})\, \oplus \, {\rm so}(V_8^{(2)})
    \, \oplus \,  V_8^{(1)} \otimes  V_8^{(2)}
\lbl{4.6}
\ee

In addition to 1-vectors and 2-vectors in ${\cal C \ell}(8)$, we
also have spinors as elements of left (right) minimal ideals
of ${\cal C \ell}(8)$. The dimension of a given minimal left ideal
of ${\cal C \ell}(8)$ is $2^{8/2} = 16$. There are 16 linearly
independent basis spinors $\zeta_{ A},~{ A} =1,2,...,16$,
and they span the spinor space $S_8$ that decomposes into
two invariant subspaces $S_8^+$ and $S_8^-$ forming,
respectively, the left-handed and right-handed spinor
representation of so(8). From the basis spinors of
${\cal C \ell}(V_8^{(1)})$ and ${\cal C \ell}(V_8^{(2)})$,
we can form the following objects that live in
${\cal C \ell}(16)={\cal C \ell}(8)\otimes {\cal C \ell}(8)$:
\be    
    \zeta_{\bar A}^{+(1)} \otimes  \zeta_{\bar B}^{+(2)}\; ,
   \qquad\qquad \zeta_{\bar A}^{-(1)} \otimes  \zeta_{\bar B}^{-(2)}
\lbl{4.7}
\ee
\be    
    \zeta_{\bar A}^{-(1)} \otimes  \zeta_{\bar B}^{+(2)}\; ,
   \qquad\qquad \zeta_{\bar A}^{+(1)} \otimes  \zeta_{\bar B}^{-(2)}
\lbl{4.8}
\ee    
where ${\bar A},~{\bar B} = 1,2,...,8$. There are
$64 + 64+64+64 = 256 = 2^{16/2}$ basis spinors of
${\cal C \ell}(16)$. Let us retain only 126 of them, say
those of eq. (\ref{4.7}) that form a positive chiral representation
(left handed) of so(16). The objects   
 $\zeta_{\bar A}^{+(1)} \otimes  \zeta_{\bar B}^{+(2)}$ and
 $\zeta_{\bar A}^{-(1)} \otimes  \zeta_{\bar B}^{-(2)}$
 span the 128-dimensional space $S_{16}^+$ of positive chiral
 spinors of ${\cal C \ell}(16)$.
 
 As it has been discussed
 in \ci{BaezOctonions,CastroE8}, the space
 \be
       {\rm so}(V_8^{(1)})\, \oplus \, {\rm so}(V_8^{(2)})
    \, \oplus \,  V_8^{(1)} \otimes  V_8^{(2)}
    \, \oplus \, 
    S_{8}^{+(1)} \otimes  S_{8}^{+(2)}
    \, \oplus \, S_{8}^{-(1)} \otimes  S_{8}^{-(2)} = {\rm e}_8
\lbl{4.9}
\ee
is the Lie algebra e$_8$ of E$_8$. Its dimension is exactly
right, namely,  $120+64+64=248$, and eq.\,(\ref{4.9}) is,
in fact, a decomposition of
\be
    {\rm e}_8 = {\rm so} (16) \oplus {\rm S}_{16}^+ 
\lbl{4.10}
\ee
where  $S_{16}^+$ forms the 128-dimensional irreducible
representation of so(16). In eq.\,(\ref{4.9}),
$V_8^{(1)}$ and $V_8^{(2)}$ are the vector representation
spaces for the two copies of so(8), whereas
$ S_{8}^{+(1)},~  S_{8}^{+(2)}$ and
$S_{8}^{-(1)},~S_{8}^{-(2)}$ are the corresponding spinor
representation spaces.

\section{Discussion and conclusion}

The exceptional group E$_8$ is somewhat enigmatic. Its smallest
representation is the 248-dimensional adjoint representation, and the
space in which $E_8$ acts as the isometry group cannot be determined
without first defining $E_8$\,\ci{BaezOctonions}.
Eqs.\,(\ref{4.9}) and (\ref{4.10}) demonstrate the curious fact
that e$_8$ consists of the Lie algebra of rotations so(16) in a
16-dimensional vector space and the corresponding spinor space
$S_{16}^+$ on which the generators of so(16) act. Both, the
generators of so(16) and the spinors of $S_{16}^+$ are
elements of ${\cal C \ell}(16)$, which contains e$_8$ as a
subspace.

${\cal C \ell}(16)$ is indeed a big space, probably too
big to be useful for the unification of the currently known
interactions and particles because it "predicts" too many
new particles and interactions. Therefore, it seems more reasonable
to stay with (\ref{4.9}), which also describes e$_8$ but with the
corresponding spaces being subspaces of ${\cal C \ell}(V_8^{(1)})$
or ${\cal C \ell}(V_8^{(2)})$. The dimension of the latter is $2^8=256$,
which is modest in comparison with the dimensionality
$2^{16} = 256\times 256$ of ${\cal C \ell}(16)$. Although
${\cal C \ell}(16)= {\cal C \ell}(8)\otimes {\cal C \ell}(8)$,
such a product is not necessary for the construction of
e$_8$ according to eq.\,(\ref{4.9}). Only the direct products
of the vector and spinor subspaces of the two copies of
${\cal C \ell}(8)$, but not of so(8), are necessary.
We have arrived at ${\cal C \ell}(8)$ quite naturally by
starting from the 4-dimensional spacetime $M_4$. From its tangent
vectors, we can generate the 16-dimensional Clifford algebra
${\cal C \ell}(1,3)$, to which we can
ascribe a physical interpretation
as the space of oriented $r$-vectors associated with strings and
branes\,\ci{Castro}--\ci{PavsicArena}, \ci{PavsicMaxwell}--\ci{AuriliaFuzzy}.
We consider ${\cal C \ell}(1,3)$ as a vector space
$V_{8,8}$ in which one can perform 120 independent rotations.
However, there are only 32 independent basis elements $\gam_A$ of
${\cal C \ell}(1,3)$, which, since they form an algebra, can
generate only 32 independent rotations in ${\cal C \ell}(1,3)$,
corresponding to left and right transformations in eq.\,(\ref{3.1}).
The fact that the $\gam_A$ form an algebra is now an undesirable
property because the product of two such gammas gives another
gamma in the same set. Thus we have concentrated on the vector
nature of $V_{8,8}$ and considered instead of $\{\gam_A \}$
another basis $\{\Gam_A \}$, whose objects do not close
an algebra and can thus form an outer product
$\Gam_{AB}=\Gam_A \wg \Gam_B$.
Next, it has turned out to be convenient to split
$V_{8,8}$ into two subspaces $V_8^{(1)}$ and $V_8^{(2)}$,
with bases $\{ \Gam_{\bar A}^{(1)} \}$ and
$\{ \Gam_{\bar A}^{(2)} \}$, ${\bar A} = 1,2,...,8$.
The bivectors $\Gam_{{\bar A}{\bar B}}^{(1)}$ of\, 
${\cal C \ell}(V_8^{(1)})$ and $\Gam_{{\bar A}{\bar B}}^{(2)}$ of 
${\cal C \ell}(V_8^{(2)})$ are generators of rotations in
the corresponding vector spaces. We have also considered
spinors of the Clifford algebras ${\cal C \ell}(V_8^{(1)})$ and
${\cal C \ell}(V_8^{(2)})$. Thus we have arrived at e$_8$ expressed
in the form (\ref{4.9}) by using only the elements of the two
copies of ${\cal C \ell}(8)$ and the direct product of the
vector and spinor representation spaces. Even if the decomposition
(\ref{4.9})
is a known result, it is fascinating to see how it fits into a straightforward
extension of 4-dimensional spacetime $M_4$ to the 16-dimensional
space of $r$-volumes. The latter space, as shown
in refs.\,\ci{Castro}--\ci{PavsicArena},\ci{PavsicMaxwell}--\ci{AuriliaFuzzy},
can be associated with strings
and branes, fundamental physical objects that, in many respects,
are very promising, although not the only ones considered in the
literature, for unification and quantum gravity.

\vs{6mm}

{\bf Acknowledgement}

This work was supported by the Ministry of High Education, Science
and Technology of Slovenia.

\end{document}